\documentclass[journal=nalefd,email=false,manuscript=letter,layout=twocolumn]{achemso}

\usepackage{graphicx,amssymb,amsfonts,amsmath,subfigure, upgreek}

\title{Tuning Interfacial Ferromagnetism in LaNiO$_3$/CaMnO$_3$ Superlattices by Stabilizing Non-Equilibrium Crystal Symmetry}

\author{C. L. Flint}
\email{cflint@stanford.edu}
\affiliation{Department of Materials Science and Engineering, Stanford University, Stanford, CA 94305, United States}
\alsoaffiliation{Geballe Laboratory for Advanced Materials, Stanford University, Stanford, CA 94305, United States}
\author{A. Vailionis}
\affiliation{Geballe Laboratory for Advanced Materials, Stanford University, Stanford, CA 94305, United States}
\author{H. Zhou}
\affiliation{Advanced Photon Source, Argonne National Laboratory, Argonne, Illinois 60439, United States, United States}
\author{H. Jang}
\author{J.-S. Lee}
\affiliation{Stanford Synchrotron Radiation Lightsource, SLAC National Accelerator Laboratory, Menlo Park, CA 94025, United States}
\author{Y. Suzuki}
\affiliation{Geballe Laboratory for Advanced Materials, Stanford University, Stanford, CA 94305, United States}
\alsoaffiliation{Department of Applied Physics, Stanford University, Stanford, CA 94305, United States}
\date{\today}
\begin{document}

\newpage

\begin{abstract}
%\begin{figure}
%	  \includegraphics{FigureToC.eps}
%   \label{fig:figToC4}
%\end{figure}

Perovskite oxide heterostructures offer an important path forward for stabilizing and controlling low-dimensional magnetism. One of the guiding design principles for these materials systems is octahedral connectivity.   In superlattices composed of perovskites with different crystal symmetries, variation of the relative ratio of the constituent layers as well as the individual layer thicknesses gives rise to non-equilibrium crystal symmetries that, in turn, lead to unprecedented control of interfacial ferromagnetism. We have found that in superlattices of CaMnO$_3$ (CMO) and LaNiO$_3$ (LNO), interfacial ferromagnetism can be modulated by a factor of three depending on LNO and CMO layer thicknesses as well as their relative ratio. Such an effect is only possible due to the non-equilibrium crystal symmetries at the interfaces and can be understood in terms of the anisotropy of the exchange interactions and modifications in the interfacial Ni-O-Mn and Mn-O-Mn bond angles and lengths with increasing LNO layer thickness. These results demonstrate the potential of engineering non-equilibrium crystal symmetries in designing ferromagnetism.
\end{abstract}

Transition metal perovskite oxides exhibit a wide range of ground states which are a manifestation of the delicate balance of the lattice, charge, and spin degrees of freedom in these materials. Competing interactions with similar energy scales mean that small perturbations, be they external fields, pressure or other parameters, can give rise to large changes in magnetic and electronic properties. In a transition metal perovskite oxide with the ABO$_3$ structure, BO$_6$ octahedra form building blocks and their relative connectivity can dramatically change its properties. In bulk single crystals, high pressure has been used to substantially modify the ground states of some of these transition metal perovksite oxides\cite{Laukhin1997}. More recently, there have been theoretical studies indicating that stabilizing new crystal symmetries via octahedra rotation patterns in oxide heterostructures may give rise to unexpected emergent behavior\cite{Rondinelli2012,He2010}.
For example, Rondinelli and Fennie have predicted ferroelectricity in cation-ordered LaGaO$_3$/YGaO$_3$ superlattices due to stabilization of unique octahedral rotation patterns\cite{Rondinelli2012b}. 

Due to their enhanced experimental signal arising from an increased number of interfaces, superlattices are model systems for exploring interfacial electronic and magnetic phenomena that are driven by octahedral connectivity. In ferromagnetic systems such as La$_{0.7}$Sr$_{0.3}$MnO$_3$/Eu$_{0.7}$Sr$_{0.3}$MnO$_3$, LaMnO$_3$/SrTiO$_3$, or La$_{0.7}$Sr$_{0.3}$MnO$_3$, experimental studies have shown that the magnetic properties are tunable through
interfacial MnO$_6$ octahedral tilt and rotation\cite{Zhai2014}\cite{Moon2014a,Moon2014b}\cite{Liao2016}. Grutter et al. have also attributed the suppression of emergent ferromagnetism in CaRuO$_3$/CaMnO$_3$ (CRO/CMO) superlattices to independent specific octahedral rotation orientations\cite{Grutter2016}. In these CRO/CMO superlattices, the relaxed strain state of the superlattices meant that the superlattice layers could re-orient independently from one another, thus modulating the ferromagnetism. By modifying octahedral connectivity, we can stabilize crystal symmetries not observed in the bulk, thereby tuning interfacial magnetism.

In this paper, we show how octahedral connectivity can be used to stabilize non-equilibrium crystal symmetries that can suppress or enhance interfacial ferromagnetism in coherently strained LaNiO$_3$/CaMnO$_3$ (LNO)$_N$/(CMO)$_M$ superlattices. We establish that non-equilibrium crystal symmetries can be stabilized in superlattices composed of constituent materials with different bulk crystal symmetries. We find that different non-equilibrium crystal symmetries can be stabilized by varying the LNO and CMO layer thicknesses. In our superlattices, the magnitude of octahedral rotations in CMO is determined by the LNO layer thickness. However, the orientation of these octahedral rotations in CMO is controlled by the CMO layer thickness. Together, these structural modifications in LNO/CMO superlattices enable control of the interfacial ferromagnetic properties over a large range of magnitudes, leading to enhanced ferromagnetism. This demonstrates that octahedral connectivity is a promising path forward for engineering interfacial ferromagnetism at the nanoscale.

To this end, we studied (LNO)$_N$/(CMO)$_M$ superlattices on 5 mm$^2$ x 0.5 mm (001) LaAlO$_3$ (LAO) single crystal substrates, where N and M are the number of LNO unit cells and CMO unit cells per superlattice period, respectively. Two sets of superlattices were grown with M equal to 4 and 8 unit cells. For each M, N was varied from 2 to 8. To maintain comparable overall thickness, M=4 superlattice periods were repeated 10 times, and M=8 superlattice periods were repeated 8 times. Films were deposited using a 248 nm KrF laser pulsed at 1 Hz with fluence of 1.3 J/cm$^{2}$. The background pressure was 60 mTorr of O$_2$ and the substrate was heated to 700 $^{\circ}$C. Unit cell growth was monitored in-situ via reflection high energy electron diffraction (RHEED), for which intensity oscillations were observed for each superlattice, indicating smooth layer-by-layer growth.
\begin{figure*}[t]
 \centering
 \includegraphics{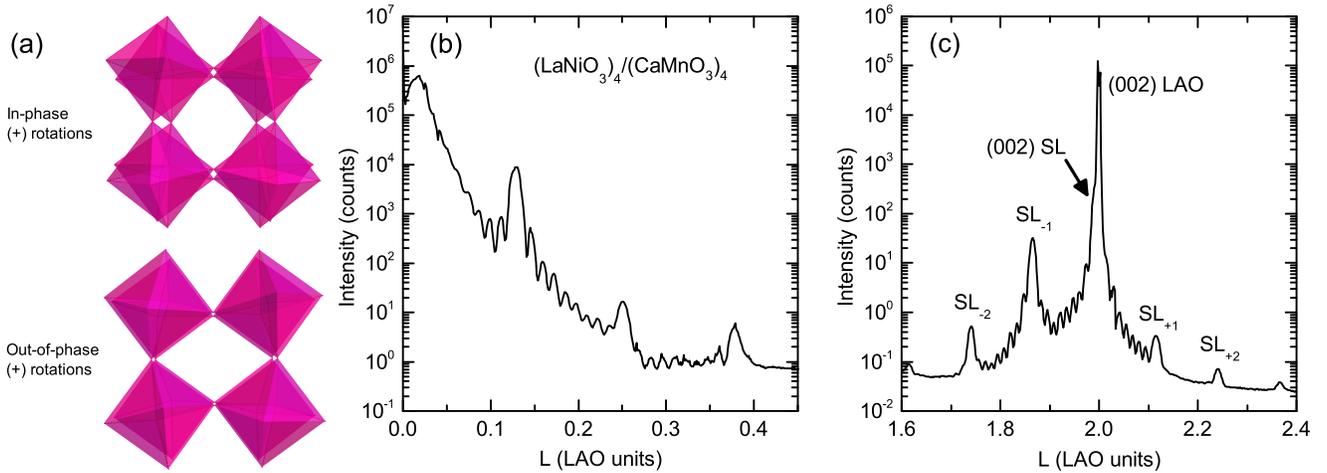}
 \caption{(a) Schematic of in-phase rotations (top) and out-of-phase rotations (bottom) present using CaMnO$_3$ as an example. The direction of the rotation axis is into the paper. (b) Specular X-ray reflectometry scan showing typical reflectivity profile of a N=4, M=4 superlattice. Determination of superlattice period via superlattice Bragg peaks is within 3$\%$ agreement with the calculated value. (c) 2$\theta$-$\theta$ scan around the (002) LAO peak. Superlattice Bragg peaks and superlattice period thickness fringes are clearly seen, indicating high structural quality.}
 \label{fig:fig1}
\end{figure*}

Structural quality was characterized ex-situ using x-ray reflectivity (XRR), x-ray diffraction (XRD), and atomic force microscopy (AFM). XRR (Figure 1b) was performed at beamline 13-3 at the Stanford Synchrotron Radiation Lightsource. 2$\uptheta$-$\uptheta$ XRD scans (Figure 1c) were performed at beamline 12ID-D at the Advanced Photon Source and indicate clear superlattice Bragg peaks and superlattice period thickness fringes. Visibility of total thickness fringes and superlattice Bragg peaks demonstrates high sample crystalline quality and layering. AFM of the superlattices reveals a surface roughness of less than half a unit cell, consistent with the smooth growth of CMO at these conditions. Therefore XRR, XRD, and AFM all confirm high quality and precise control of the superlattice growth in this study.

Reciprocal space maps of the (103) diffraction peaks reveal that both the CMO and LNO layers are coherently strained to the underlying LAO substrates in all of our superlattices. It is important to note that LAO forms a rhombohedral crystal lattice in the bulk with a pseudocubic lattice parameter of a=3.798 \AA. LNO also has a rhombohedral unit cell that can be approximated by a pseudocubic lattice parameter of a=3.85 \AA \cite{Sreedhar1992}. CMO has an orthorhombic unit cell that can be approximated by a pseudocubic lattice parameter of a=3.73 \AA \cite{Paszkowicz2010}. In perovskite oxides, octahedral rotations are largely responsible for the various crystal symmetries that exist between compounds. For example, rhombohedral LNO has \textit{a$^-$a$^-$a$^-$} rotations, using Glazer notation\cite{Glazer1975}. In this notation $^-$ refers to out-of-phase rotations while $^+$ refers to in-phase rotations. On the other hand, CMO has \textit{a$^-$a$^-$c$^+$} rotations with in-phase rotations along the \textit{c} direction. Figure 1a illustrates the in-phase and out-of-phase rotations of these oxygen octahedra. Coherent strain therefore may impose a non-equilibrium unit cell and non-equilibirium octahedral rotations in the CMO and LNO layers depending on their relative thicknesses. 

To probe how coherent strain modifies the CMO and LNO atomic structures in the superlattices, we examined half-order x-ray diffraction peaks at beamline 12ID-D of the Advanced Photon Source at Argonne National Laboratory. For perovskite oxides, differences in bond angles, bond lengths and crystal symmetries can be described in terms of how the oxygen octahedra are rotated and tilted relative to one another; this is sometimes referred to as octahedral connectivity. This connectivity can be analyzed in terms of the existence and intensities of half-order diffraction peaks\cite{Glazer1975}. These diffraction peaks (Figure 2), and therefore the structural accommodations, are distinctly different for the M=4 and M=8 superlattices, thereby affecting the interfacial ferromagnetism in different ways. 

\begin{figure}[t]
 \centering
 \includegraphics{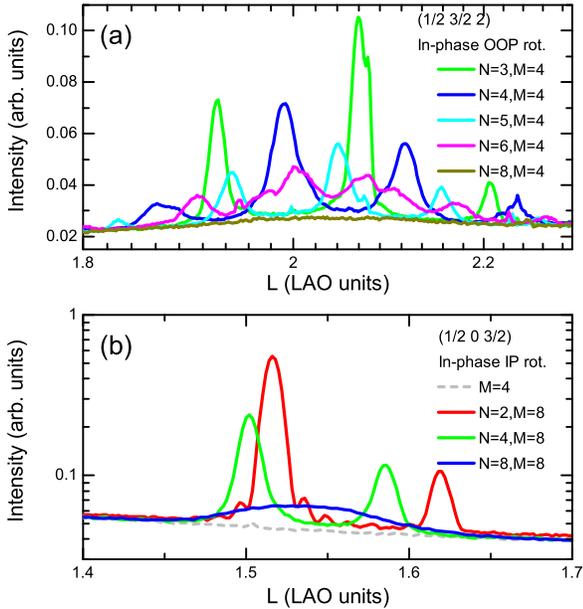}
 \caption{(a) X-ray diffractogram of ($\frac{1}{2}$ $\frac{3}{2}$ 2) half-order Bragg peak due to out-of-plane, in-phase rotations in CMO=4 u.c. superlattices. As LNO layer thickness N increases, in-phase CMO rotations decrease and eventually disappear. (b) X-ray diffractogram of ($\frac{1}{2}$ 0 $\frac{3}{2}$) half-order Bragg peak due to in-plane, in-phase rotations in CMO=8 u.c. superlattices. As LNO layer thickness N increases, in-phase CMO rotations decrease and nearly disappear by N=8. M=4 superlattice (dashed line) is included to show lack of in-phase IP rotations in M=4 superlattices.}
 \label{fig:fig2}
\end{figure}

For M=4 superlattices, Figure 2a presents the evolution of the ($\frac{1}{2}$ $\frac{3}{2}$ 2) half-order diffraction peak, which corresponds to the \textit{c$^+$}-type, out-of-plane, in-phase rotation in CMO. As the LNO thickness increases, the intensity of the in-phase rotations is reduced. It eventually disappears entirely by N=8. We also found throughout all M=4 samples that there are no in-plane, in-phase rotations associated with the (0 $\frac{1}{2}$ $\frac{3}{2}$) and ($\frac{1}{2}$ 0 $\frac{3}{2}$) peaks (dashed line in Figure 2b). Thus in the thin CMO regime, the growth axis is the preferred in-phase axis. 

From this data, we can conclude that increasing the LNO layer thickness diminishes the out-of-plane, in-phase rotations in CMO, possibly imposing the LNO \textit{a$^-$a$^-$a$^-$} out-of-phase rotation pattern throughout the LNO and CMO layers of the superlattice. Unlike for in-phase rotations, there is no unique out-of-phase rotation half-order diffraction peak, and LAO and LNO both exhibit out-of-phase rotations. Therefore we cannot attribute a single peak intensity to \textit{a$^-$a$^-$a$^-$} rotations in CMO using x-ray diffraction. As a result, it is not possible to definitively determine whether the LNO rhombohedral \textit{a$^-$a$^-$a$^-$} symmetry is established in the CMO or whether the CMO simply loses its in-phase rotations, resulting in \textit{a$^-$a$^-$c$^0$} rotations . At a minimum the in-phase rotations have been unrotated. 

Changes in rotation pattern are accommodated via changes in Mn-O-Mn bond length as well as angle. As the out-of-plane, in-phase CMO rotations unrotate, the in-plane bond angles straighten. As a result, the Mn-Mn distance is now larger, which increases the unit cell spacing. However, as confirmed via reciprocal space mapping, these superlattices are coherently strained to the substrate. Therefore the in-plane lattice constant is fixed. Hence, straightening of the CMO in-plane bond angles must be accompanied by a corresponding shortening of the in-plane Mn-O bond lengths. These modifications to the Mn-O bond are expected to have significant consequences for the exchange interactions at the interface. 

For M=8 superlattices, we do not observe peaks at the ($\frac{1}{2}$ $\frac{3}{2}$ 2) half-order diffraction index. Therefore, unlike M=4 superlattices, M=8 superlattices do not possess out-of-plane, in-phase rotations. By investigating (0 $\frac{1}{2}$ $\frac{3}{2}$) and ($\frac{1}{2}$ 0 $\frac{3}{2}$) type peaks, we find that the in-phase rotation axis of M=8 superlattices is oriented in-plane, with equal preference for the (1 0 0) ((0 $\frac{1}{2}$ $\frac{3}{2}$) half order peak) and (0 1 0) (($\frac{1}{2}$ 0 $\frac{3}{2}$) half order peak) axes. This finding is consistent with the preferred orthorhombic growth direction observed in manganite thin films\cite{Vailionis2011} and suggests that the stabilization of out-of-plane in-phase orientation for M=4 superlattices may be a finite size effect in the ultra-thin regime. The evolution of the ($\frac{1}{2}$ 0 $\frac{3}{2}$) peak as a function of LNO layer thickness is shown in Figure 2b. For M=8 superlattices, even though the CMO in-phase rotations are oriented in-plane instead of out-of-plane, increasing N has the same effect of straightening the in-phase rotations. By N=8, M=8, the in-phase rotations nearly have disappeared. 

\begin{figure}
	  \includegraphics{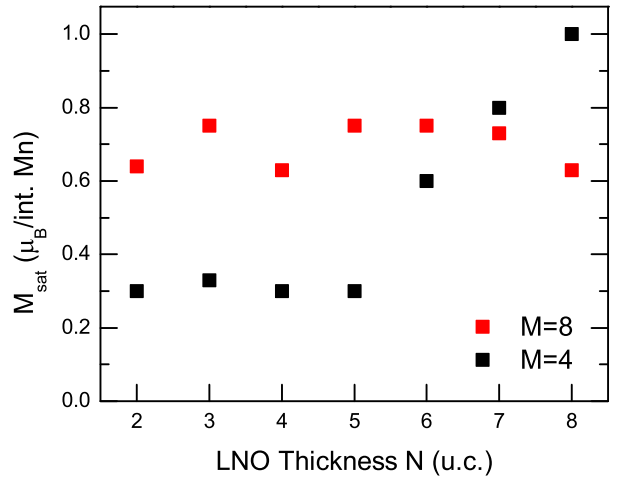}
\caption{LNO layer thickness dependence of (LNO)$_N$/(CMO)$_M$ superlattice saturated magnetic moment at 7 T and 10 K. (LNO)$_N$/(CMO)$_4$ superlattices with M=4 exhibit a nearly constant M$_{sat}$ at low N and increasing M$_{sat}$ with N$>$5. Superlattices with M=8 exhibit nearly a constant M$_{sat}$ across the full range of LNO thicknesses.}
    \label{fig:fig3}
\end{figure}

Bulk magnetization measurements revealed ferromagnetic signal for all superlattices. Samples were measured at 10 K in fields up to 7 T using a SQUID magnetometer. Saturated magnetic moments for each superlattice are summarized in Figure 3. Due to small amounts of paramagnetic and ferromagnetic contamination of the substrates\cite{Khalid2010}, a background substraction was performed to determine the superlattice contribution to the magnetization. The saturated magnetic moment has been normalized to the number of interfacial Mn ions for comparison with previous work on CMO-based superlattices\cite{Grutter2016,Takahashi2001,Freeland2010,He2012,Grutter2013}. Interfacial ferromagnetism in LNO/CMO has been explained by a double-exchange based model of interfacial ferromagnetism where a small amount of electrons from the metallic LNO layer leak into the interfacial CMO layer and induce ferromagnetism\cite{Nanda2007}. In this scenario, the CMO layer determines the ferromagnetic properties via Mn$^{4+}$-Mn$^{3+}$ double exchange\cite{Nanda2007}. 

	However, there are two features in our M=4 and M=8 samples that are unexplained by this model alone: (1) at low N (N$\leq$4), M=8 superlattices have approximately double the saturated magnetic moment of the M=4 superlattices and (2) at large N (N$>$5), the saturated magnetic moment of M=8 superlattices is constant, while the saturated magnetic moment of the M=4 superlattices strongly depends on the LNO layer thicknesses.

\begin{figure}
	  \includegraphics{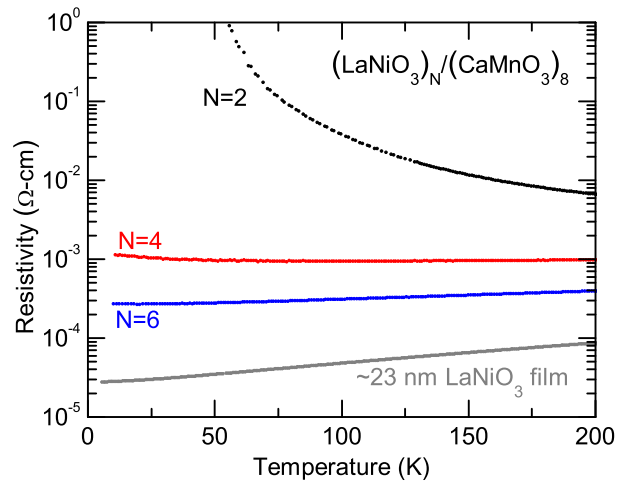}
\caption{Temperature dependence from 10-200 K of superlattice resistivity for M=8 N=2,4,6 superlattices. Included is temperature dependence from 5-200 K of LNO thin film resistivity for comparison. Metal-insulator transition at N=4 and gradual approach to bulk LNO value are observed, consistent with previous results\cite{Grutter2013,Scherwitzl2011}.}
    \label{fig:fig4}
\end{figure}

Since the interfacial double exchange model depends on electronic properties of the LNO layer, we performed electronic transport measurements to characterize the superlattice conductivity. Figure 4 shows resistivity versus temperature of M=8 N=2, 4, 6 superlattices from 5--200 K. A 23 nm thick film of LNO grown under the same conditions is provided for comparison. M=4 superlattices are omitted for clarity, but show a similar trend. For M=4 and M=8 superlattices, there is a metal-insulator transition at N=4 unit cells, with N$\geq$4 superlattices being metallic. While all superlattices with N$<$4 are insulating, they are still magnetic. This means that at low N, an additional interfacial ferromagnetic mechanism must be operative. The most likely mechanism is a Mn-O-Ni superexchange interaction that we have described in more detail elsewhere.\cite{Flint2017}\bibnote{It should be noted that in a previous study, no ferromagnetism was observed in M=8 insulating superlattices\cite{Grutter2013} A careful comparison of the transport data of the M=8 superlattices in this current study to those of the previous study reveals that the present SLs have higher resistivity at a given N and M. These differences in resistivity suggest there may be differences in oxygen stoichiometry and cation oxidation states at the interface between sample sets. These differences are plausible given the different growth conditions}. Since the M=4 and M=8 superlattices have similar resistivity behavior, the transport data does not explain the difference in magnetic moment between M=4 and M=8 superlattices at low N nor does it explain the difference in magnetic moment trends as a function of LNO layer thickness. 

\begin{figure*}[t]
 \centering
  \includegraphics{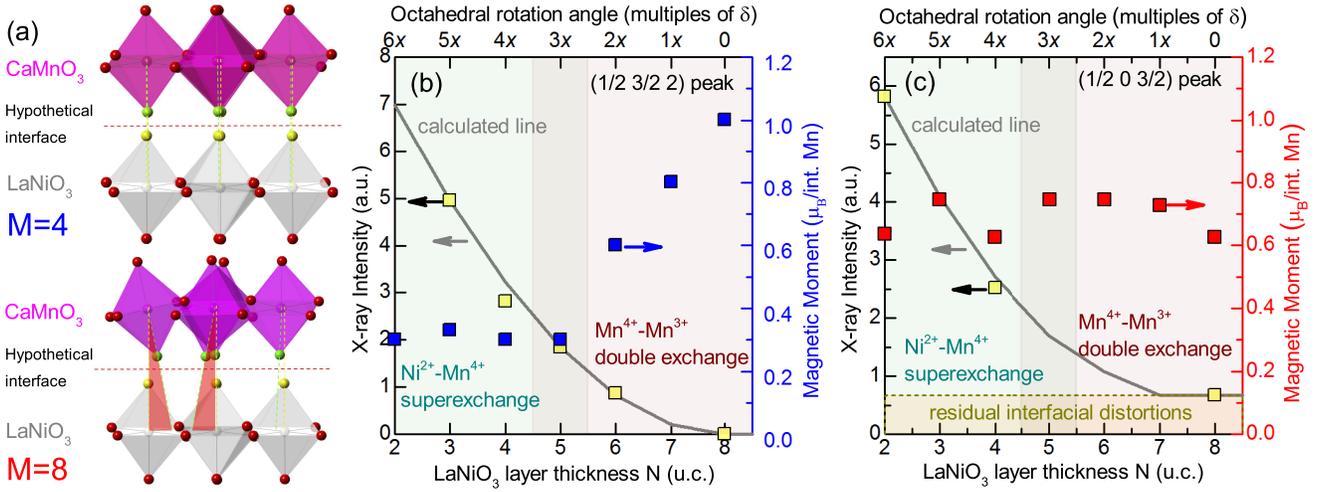}
 \caption{(a) Hypothetical interfacial alignment for M=4 and M=8 superlattices demonstrating M=8 mismatch due to in-plane orthorhombic orientation. (b) X-ray intensity of ($\frac{1}{2}$ $\frac{3}{2}$ 2) half-order Bragg peak due to out-of-plane, in-phase rotations in CMO=4 u.c. superlattices (left axis). As LNO layer thickness N increases, in-phase CMO rotations decrease and eventually disappear. The calculated line shows expected intensity for a constant decrease in the rotation angle with increasing LNO layer thickness. Saturated magnetic moment (right axis) increases once the double-exchange interaction is dominant. (b) X-ray intensity of ($\frac{1}{2}$ 0 $\frac{3}{2}$) half-order Bragg peak due to in-plane, in-phase rotations in CMO=8 u.c. superlattices (left axis). As LNO layer thickness N increases, in-phase CMO rotations decrease and nearly disappear by N=8. The calculated line shows the experimental data is well fit to the same model as M=4, with the addition of a constant intensity offset due to interfacial mismatch.}
 \label{fig:fig5}
\end{figure*}

We must therefore turn to alternative explanations for the observed ferromagnetism. Given the evolution of the structural data as a function of N for M=4, a closer look at the relationship between structural and magnetic properties in these superlattices is warranted. We propose a model based on tuning octahedral rotations that depends on the interfacial alignment between LNO and CMO. Figure 5a illustrates this alignment and the differences between M=4 and M=8 superlattices. In addition to the magnetization data, Figures 5b,c depict the x-ray intensity of the in-phase rotation peaks with a calculated line that assumes a constant change ($\delta$) in the in-phase rotation angle for each additional LNO unit cell added to the superlattice. We now discuss the correlation between this x-ray data and the magnetization data in more depth.

For M=4 superlattices, as the thickness of the LNO metallic layer is increased from N=5 to N=8, the saturated moment per interfacial Mn shows a drastic increase (i.e., more than triples). This increase is not sufficiently explained within the conventional model of interfacial itinerant electron-based double exchange interaction due to the adjacent metallic layer. As mentioned previously, it is known for perovskite oxides that changes in octahedral rotations modify the M-O-M bond angles and bond lengths and that these effects can impact the magnetic properties. Within this context, a possible explanation for the observed magnetic trend is enhancement of the interfacial double exchange mechanism as a result of the stabilization of non-equilibrium crystal symmetries. The modification of the CMO symmetry to reduce the orthorhombic distortion may enhance the interfacial double exchange mechanism. One possible reason for this modification is the influence of biaxial strain from the LAO substrate, which leads to a monoclinic distortion in bulk LNO with out-of-plane rotations that are much larger than the in-plane rotations\cite{May2010}. On the other hand, epitaxially strained CMO on LAO is predicted to have large in-plane and out-of-plane rotations \cite{Aschauer2013}. As the LNO thickness increases and the CMO adopts the \textit{a$^-$a$^-$c$^0$} pattern, the reduction of in-plane rotation angles in CMO would increase the Mn-O-Mn bond angles and improve the double-exchange interaction between Mn$^{4+}$-Mn$^{3+}$ ions. This symmetry change in the CMO layer can be easily accommodated across the interface because changes to the out-of-plane rotations affect the in-plane rotation angles. 

In addition to explaining the trend of increasing magnetization at N$>$5, the symmetry change from \textit{a$^-$a$^-$c$^+$} to \textit{a$^-$a$^-$c$^0$} is also consistent with constant magnetization in N=2-5 superlattices. Changes to the out-of-plane rotations from increasing LNO layer thickness do not strongly influence the apical oxygens across the LNO--CMO interface. Only rotations angles perpendicular to the rotation axes (i.e. in-plane rotation angles) are affected. Since the rotations along the out-of-plane axis are able to freely rotate, leaving the apical oxygens undisturbed, the dominant mechanism at low N---Ni-O-Mn superexchange  across the interface---is unaffected.  

However, turning to the M=8 superlattices, one observes that the LNO layer thickness has little influence on the magnetic moment. In other words, even though the LNO layer leads to a similar reduction in the CMO in-phase rotation, it does not result in a similar increase in magnetic moment. This suggests that while the crystal symmetry control via LNO layer thickness is important for determining magnetic properties, it depends critically on the CMO orthorhombic orientation, which is determined by the CMO layer thickness. In M=8 superlattices, the CMO orthorhombic axis is in-plane, which means reductions in the in-phase rotations should directly affect the interfacial apical oxygens. In these superlattices, one may expect changes in the in-phase rotations in CMO to modify the Ni-O-Mn bond angle. However, since no change is observed in the saturated magnetic moment of these samples, the superexchange and double-exchange mechanisms must be unaffected. Thus, the interface Ni-O-Mn and Mn-O-Mn bond angles are similarly unaffected. 

One possible scenario then, is that the CMO interface maintains a constant and distinct rotation pattern from the interior of the CMO layer. As the majority of the CMO unrotates with increasing LNO layer thickness, the interface maintains its structural state. For the x-ray intensity in Figure 5c, this would be the equivalent to a constant offset in the modulation of the x-ray intensity with LNO thickness. Indeed, from Figure 5c we find that the x-ray data for M=8 superlattices matches well with this model. These results suggest that in the M=8 superlattices, the interface may adopt a distinct structural state, separate from the CMO and LNO rotation patterns. This intermediate interfacial state, arising from the in-plane CMO orthrohombic orientation (M=8 superlattices), results in lower tunability of the ferromagnetism within this LNO thickness range compared to that arising from the out-of-plane CMO orthorhombic orientation (M=4 superlattices). However, this interfacial state in M=8 superlatices also leads to a higher saturated magnetic moment at low N. 

The dependence of the magnetic moment evolution on the orientation of the CMO in-phase rotation axis (out-of-plane for M=4 and in-plane for M=8) suggests that the transition from \textit{a$^-$a$^-$c$^+$} to \textit{a$^-$a$^-$c$^0$} may be accommodated differently than the transition from \textit{a$^-$c$^+$a$^-$} or \textit{c$^+$a$^-$a$^-$} to \textit{a$^-$c$^0$a$^-$} or \textit{c$^0$a$^-$a$^-$}, respectively. Further studies are needed to understand exactly how the interface accommodates the transition from CMO-type rotations to LNO-type rotations. One critical tuning parameter may be the N/M ratio. For M=4 superlattices, the increase in magnetization is not observed until N/M=3/2, while for M=8 superlattices, we only investigated up to N/M=1. In this thickness regime, while Figure 2b demonstrates that the intensity of the CMO in-phase rotation in M=8 superlattices is nearly diminished by N=8, a small, broad peak is still apparent. It is clear from comparison of the XRD intensities that the M=8 CMO in-phase rotations are much more strongly diminished by N/M=1 than those in the M=4 case. However, this remnant intensity supports the assertion that the CMO rotations for M=8 and N=8 are in some intermediate state due to difficulty in accommodating changes in phase and magnitude of the in-plane \textit{c$^+$} rotation. 

By investigating superlattices with 4 and 8 u.c. of CMO across a range of LNO layer thickness, we have demonstrated that the stabilization of non-equilibrium crystal symmetries of a material via heteorepitaxy can give rise to a range of interfacial ferromagnetic responses via octahedral connectivity. We find that the LNO thickness controls the magnitude of the CMO in-phase rotations, and the CMO thickness determines the rotation orientation. Moreover, LNO layer thicknesses approaching 8 u.c. suppress the orthorhombic symmetry of the CMO layers. Our studies indicate that differences in the emergent ferromagnetic behavior of superlattices with 4 and 8 u.c. of CMO is the result of how the anisotropic octahedral rotations influence the strength of the anisotropic ferromagnetic exchange interactions at the LNO-CMO interface and demonstrates the complex interplay of in-phase and out-of-phase rotations on the functional properties. This understanding of the relationship between crystal symmetry and interfacial ferromagnetism is important for the future development of oxide based electronics and spintronics. 

\begin{acknowledgement}
Research was supported by the U.S. Department of Energy, Director, Office of Science, Office of Basic Energy Sciences, Division of Materials Sciences and Engineering under Contract No. DESC0008505. Use of the Stanford Synchrotron Radiation Light source, SLAC National Accelerator Laboratory, is supported by the U.S. Department of Energy, Office of Science, Office of Basic Energy Sciences under Contract No. DE-AC02-76SF00515. The Advanced Light Source is supported by the Director, Office of Science, Office of Basic Energy Sciences, of the U.S. Department of Energy under Contract No. DE-AC02-05CH11231. Use of the Advanced Photon Source was supported by the U. S. Department of Energy, Office of Science, Office of Basic Energy Sciences, under Contract No. DE-AC02-06CH11357.
\end{acknowledgement}

\bibliography{main}

\providecommand{\latin}[1]{#1}
\makeatletter
\providecommand{\doi}
  {\begingroup\let\do\@makeother\dospecials
  \catcode`\{=1 \catcode`\}=2\doi@aux}
\providecommand{\doi@aux}[1]{\endgroup\texttt{#1}}
\makeatother
\providecommand*\mcitethebibliography{\thebibliography}
\csname @ifundefined\endcsname{endmcitethebibliography}
  {\let\endmcitethebibliography\endthebibliography}{}
\begin{mcitethebibliography}{25}
\providecommand*\natexlab[1]{#1}
\providecommand*\mciteSetBstSublistMode[1]{}
\providecommand*\mciteSetBstMaxWidthForm[2]{}
\providecommand*\mciteBstWouldAddEndPuncttrue
  {\def\EndOfBibitem{\unskip.}}
\providecommand*\mciteBstWouldAddEndPunctfalse
  {\let\EndOfBibitem\relax}
\providecommand*\mciteSetBstMidEndSepPunct[3]{}
\providecommand*\mciteSetBstSublistLabelBeginEnd[3]{}
\providecommand*\EndOfBibitem{}
\mciteSetBstSublistMode{f}
\mciteSetBstMaxWidthForm{subitem}{(\alph{mcitesubitemcount})}
\mciteSetBstSublistLabelBeginEnd
  {\mcitemaxwidthsubitemform\space}
  {\relax}
  {\relax}

\bibitem[Laukhin \latin{et~al.}(1997)Laukhin, Fontcuberta, Garc\'{\i}a-Mu\~noz,
  and Obradors]{Laukhin1997}
Laukhin,~V.; Fontcuberta,~J.; Garc\'{\i}a-Mu\~noz,~J.~L.; Obradors,~X.
  \emph{Phys. Rev. B} \textbf{1997}, \emph{56}, R10009--R10012\relax
\mciteBstWouldAddEndPuncttrue
\mciteSetBstMidEndSepPunct{\mcitedefaultmidpunct}
{\mcitedefaultendpunct}{\mcitedefaultseppunct}\relax
\EndOfBibitem
\bibitem[Rondinelli \latin{et~al.}(2012)Rondinelli, May, and
  Freeland]{Rondinelli2012}
Rondinelli,~J.~M.; May,~S.~J.; Freeland,~J.~W. \emph{MRS Bulletin}
  \textbf{2012}, \emph{37}, 261–270\relax
\mciteBstWouldAddEndPuncttrue
\mciteSetBstMidEndSepPunct{\mcitedefaultmidpunct}
{\mcitedefaultendpunct}{\mcitedefaultseppunct}\relax
\EndOfBibitem
\bibitem[He \latin{et~al.}(2010)He, Borisevich, Kalinin, Pennycook, and
  Pantelides]{He2010}
He,~J.; Borisevich,~A.; Kalinin,~S.~V.; Pennycook,~S.~J.; Pantelides,~S.~T.
  \emph{Phys. Rev. Lett.} \textbf{2010}, \emph{105}, 227203\relax
\mciteBstWouldAddEndPuncttrue
\mciteSetBstMidEndSepPunct{\mcitedefaultmidpunct}
{\mcitedefaultendpunct}{\mcitedefaultseppunct}\relax
\EndOfBibitem
\bibitem[Rondinelli and Fennie(2012)Rondinelli, and Fennie]{Rondinelli2012b}
Rondinelli,~J.~M.; Fennie,~C.~J. \emph{Advanced Materials} \textbf{2012},
  \emph{24}, 1918--1918\relax
\mciteBstWouldAddEndPuncttrue
\mciteSetBstMidEndSepPunct{\mcitedefaultmidpunct}
{\mcitedefaultendpunct}{\mcitedefaultseppunct}\relax
\EndOfBibitem
\bibitem[Zhai \latin{et~al.}()Zhai, Cheng, Liu, Schlep{\"u}tz, Dong, Li, Zhang,
  Chu, Zheng, Zhang, Zhao, Hong, Bhattacharya, Eckstein, and Zeng]{Zhai2014}
Zhai,~X.; Cheng,~L.; Liu,~Y.; Schlep{\"u}tz,~C.~M.; Dong,~S.; Li,~H.;
  Zhang,~X.; Chu,~S.; Zheng,~L.; Zhang,~J.; Zhao,~A.; Hong,~H.;
  Bhattacharya,~A.; Eckstein,~J.~N.; Zeng,~C. \emph{Nature Communications}
  \emph{5}, 4283 EP --\relax
\mciteBstWouldAddEndPuncttrue
\mciteSetBstMidEndSepPunct{\mcitedefaultmidpunct}
{\mcitedefaultendpunct}{\mcitedefaultseppunct}\relax
\EndOfBibitem
\bibitem[Moon \latin{et~al.}()Moon, Colby, Wang, Karapetrova, Schlep{\"u}tz,
  Fitzsimmons, and May]{Moon2014a}
Moon,~E.~J.; Colby,~R.; Wang,~Q.; Karapetrova,~E.; Schlep{\"u}tz,~C.~M.;
  Fitzsimmons,~M.~R.; May,~S.~J. \emph{Nature Communications} \emph{5}, 5710 EP
  --\relax
\mciteBstWouldAddEndPuncttrue
\mciteSetBstMidEndSepPunct{\mcitedefaultmidpunct}
{\mcitedefaultendpunct}{\mcitedefaultseppunct}\relax
\EndOfBibitem
\bibitem[Moon \latin{et~al.}(2014)Moon, Balachandran, Kirby, Keavney,
  Sichel-Tissot, Schlepütz, Karapetrova, Cheng, Rondinelli, and
  May]{Moon2014b}
Moon,~E.~J.; Balachandran,~P.~V.; Kirby,~B.~J.; Keavney,~D.~J.;
  Sichel-Tissot,~R.~J.; Schlepütz,~C.~M.; Karapetrova,~E.; Cheng,~X.~M.;
  Rondinelli,~J.~M.; May,~S.~J. \emph{Nano Letters} \textbf{2014}, \emph{14},
  2509--2514, PMID: 24697503\relax
\mciteBstWouldAddEndPuncttrue
\mciteSetBstMidEndSepPunct{\mcitedefaultmidpunct}
{\mcitedefaultendpunct}{\mcitedefaultseppunct}\relax
\EndOfBibitem
\bibitem[Liao \latin{et~al.}(2016)Liao, Huijben, Zhong, Gauquelin, Macke,
  Green, Van~Aert, Verbeeck, Van~Tendeloo, Held, Sawatzky, Koster, and
  Rijnders]{Liao2016}
Liao,~Z.; Huijben,~M.; Zhong,~Z.; Gauquelin,~N.; Macke,~S.; Green,~R.;
  Van~Aert,~S.; Verbeeck,~J.; Van~Tendeloo,~G.; Held,~K.; Sawatzky,~G.;
  Koster,~G.; Rijnders,~G. \emph{Nature Materials} \textbf{2016}, \emph{15},
  425--431\relax
\mciteBstWouldAddEndPuncttrue
\mciteSetBstMidEndSepPunct{\mcitedefaultmidpunct}
{\mcitedefaultendpunct}{\mcitedefaultseppunct}\relax
\EndOfBibitem
\bibitem[Grutter \latin{et~al.}(2016)Grutter, Vailionis, Borchers, Kirby,
  Flint, He, Arenholz, and Suzuki]{Grutter2016}
Grutter,~A.~J.; Vailionis,~A.; Borchers,~J.~A.; Kirby,~B.~J.; Flint,~C.~L.;
  He,~C.; Arenholz,~E.; Suzuki,~Y. \emph{Nano Letters} \textbf{2016},
  \emph{16}, 5647--5651, PMID: 27472285\relax
\mciteBstWouldAddEndPuncttrue
\mciteSetBstMidEndSepPunct{\mcitedefaultmidpunct}
{\mcitedefaultendpunct}{\mcitedefaultseppunct}\relax
\EndOfBibitem
\bibitem[Sreedhar \latin{et~al.}(1992)Sreedhar, Honig, Darwin, McElfresh,
  Shand, Xu, Crooker, and Spalek]{Sreedhar1992}
Sreedhar,~K.; Honig,~J.~M.; Darwin,~M.; McElfresh,~M.; Shand,~P.~M.; Xu,~J.;
  Crooker,~B.~C.; Spalek,~J. \emph{Phys. Rev. B} \textbf{1992}, \emph{46},
  6382--6386\relax
\mciteBstWouldAddEndPuncttrue
\mciteSetBstMidEndSepPunct{\mcitedefaultmidpunct}
{\mcitedefaultendpunct}{\mcitedefaultseppunct}\relax
\EndOfBibitem
\bibitem[Paszkowicz \latin{et~al.}(2010)Paszkowicz, Piętosa, Woodley,
  Dłużewski, Kozłowski, and Martin]{Paszkowicz2010}
Paszkowicz,~W.; Piętosa,~J.; Woodley,~S.~M.; Dłużewski,~P.~A.;
  Kozłowski,~M.; Martin,~C. \emph{Powder Diffraction} \textbf{2010},
  \emph{25}, 46–59\relax
\mciteBstWouldAddEndPuncttrue
\mciteSetBstMidEndSepPunct{\mcitedefaultmidpunct}
{\mcitedefaultendpunct}{\mcitedefaultseppunct}\relax
\EndOfBibitem
\bibitem[Glazer(1975)]{Glazer1975}
Glazer,~A.~M. \emph{Acta Crystallographica Section A} \textbf{1975}, \emph{31},
  756--762\relax
\mciteBstWouldAddEndPuncttrue
\mciteSetBstMidEndSepPunct{\mcitedefaultmidpunct}
{\mcitedefaultendpunct}{\mcitedefaultseppunct}\relax
\EndOfBibitem
\bibitem[Vailionis \latin{et~al.}(2011)Vailionis, Boschker, Siemons, Houwman,
  Blank, Rijnders, and Koster]{Vailionis2011}
Vailionis,~A.; Boschker,~H.; Siemons,~W.; Houwman,~E.~P.; Blank,~D. H.~A.;
  Rijnders,~G.; Koster,~G. \emph{Phys. Rev. B} \textbf{2011}, \emph{83},
  064101\relax
\mciteBstWouldAddEndPuncttrue
\mciteSetBstMidEndSepPunct{\mcitedefaultmidpunct}
{\mcitedefaultendpunct}{\mcitedefaultseppunct}\relax
\EndOfBibitem
\bibitem[Khalid \latin{et~al.}(2010)Khalid, Setzer, Ziese, Esquinazi, Spemann,
  P\"oppl, and Goering]{Khalid2010}
Khalid,~M.; Setzer,~A.; Ziese,~M.; Esquinazi,~P.; Spemann,~D.; P\"oppl,~A.;
  Goering,~E. \emph{Phys. Rev. B} \textbf{2010}, \emph{81}, 214414\relax
\mciteBstWouldAddEndPuncttrue
\mciteSetBstMidEndSepPunct{\mcitedefaultmidpunct}
{\mcitedefaultendpunct}{\mcitedefaultseppunct}\relax
\EndOfBibitem
\bibitem[Takahashi \latin{et~al.}(2001)Takahashi, Kawasaki, and
  Tokura]{Takahashi2001}
Takahashi,~K.~S.; Kawasaki,~M.; Tokura,~Y. \emph{Applied Physics Letters}
  \textbf{2001}, \emph{79}, 1324--1326\relax
\mciteBstWouldAddEndPuncttrue
\mciteSetBstMidEndSepPunct{\mcitedefaultmidpunct}
{\mcitedefaultendpunct}{\mcitedefaultseppunct}\relax
\EndOfBibitem
\bibitem[Freeland \latin{et~al.}(2010)Freeland, Chakhalian, Boris, Tonnerre,
  Kavich, Yordanov, Grenier, Zschack, Karapetrova, Popovich, Lee, and
  Keimer]{Freeland2010}
Freeland,~J.~W.; Chakhalian,~J.; Boris,~A.~V.; Tonnerre,~J.-M.; Kavich,~J.~J.;
  Yordanov,~P.; Grenier,~S.; Zschack,~P.; Karapetrova,~E.; Popovich,~P.;
  Lee,~H.~N.; Keimer,~B. \emph{Phys. Rev. B} \textbf{2010}, \emph{81},
  094414\relax
\mciteBstWouldAddEndPuncttrue
\mciteSetBstMidEndSepPunct{\mcitedefaultmidpunct}
{\mcitedefaultendpunct}{\mcitedefaultseppunct}\relax
\EndOfBibitem
\bibitem[He \latin{et~al.}(2012)He, Grutter, Gu, Browning, Takamura, Kirby,
  Borchers, Kim, Fitzsimmons, Zhai, Mehta, Wong, and Suzuki]{He2012}
He,~C.; Grutter,~A.~J.; Gu,~M.; Browning,~N.~D.; Takamura,~Y.; Kirby,~B.~J.;
  Borchers,~J.~A.; Kim,~J.~W.; Fitzsimmons,~M.~R.; Zhai,~X.; Mehta,~V.~V.;
  Wong,~F.~J.; Suzuki,~Y. \emph{Phys. Rev. Lett.} \textbf{2012}, \emph{109},
  197202\relax
\mciteBstWouldAddEndPuncttrue
\mciteSetBstMidEndSepPunct{\mcitedefaultmidpunct}
{\mcitedefaultendpunct}{\mcitedefaultseppunct}\relax
\EndOfBibitem
\bibitem[Grutter \latin{et~al.}(2013)Grutter, Yang, Kirby, Fitzsimmons, Aguiar,
  Browning, Jenkins, Arenholz, Mehta, Alaan, and Suzuki]{Grutter2013}
Grutter,~A.~J.; Yang,~H.; Kirby,~B.~J.; Fitzsimmons,~M.~R.; Aguiar,~J.~A.;
  Browning,~N.~D.; Jenkins,~C.~A.; Arenholz,~E.; Mehta,~V.~V.; Alaan,~U.~S.;
  Suzuki,~Y. \emph{Phys. Rev. Lett.} \textbf{2013}, \emph{111}, 087202\relax
\mciteBstWouldAddEndPuncttrue
\mciteSetBstMidEndSepPunct{\mcitedefaultmidpunct}
{\mcitedefaultendpunct}{\mcitedefaultseppunct}\relax
\EndOfBibitem
\bibitem[Nanda \latin{et~al.}(2007)Nanda, Satpathy, and Springborg]{Nanda2007}
Nanda,~B. R.~K.; Satpathy,~S.; Springborg,~M.~S. \emph{Phys. Rev. Lett.}
  \textbf{2007}, \emph{98}, 216804\relax
\mciteBstWouldAddEndPuncttrue
\mciteSetBstMidEndSepPunct{\mcitedefaultmidpunct}
{\mcitedefaultendpunct}{\mcitedefaultseppunct}\relax
\EndOfBibitem
\bibitem[Scherwitzl \latin{et~al.}(2011)Scherwitzl, Gariglio, Gabay, Zubko,
  Gibert, and Triscone]{Scherwitzl2011}
Scherwitzl,~R.; Gariglio,~S.; Gabay,~M.; Zubko,~P.; Gibert,~M.; Triscone,~J.-M.
  \emph{Phys. Rev. Lett.} \textbf{2011}, \emph{106}, 246403\relax
\mciteBstWouldAddEndPuncttrue
\mciteSetBstMidEndSepPunct{\mcitedefaultmidpunct}
{\mcitedefaultendpunct}{\mcitedefaultseppunct}\relax
\EndOfBibitem
\bibitem[Flint \latin{et~al.}()Flint, Jang, Lee, N'Diaye, Shafer, Arenholz, and
  Suzuki]{Flint2017}
Flint,~C.; Jang,~H.; Lee,~J.-S.; N'Diaye,~A.~T.; Shafer,~P.; Arenholz,~E.;
  Suzuki,~Y. \emph{unpublished} \relax
\mciteBstWouldAddEndPunctfalse
\mciteSetBstMidEndSepPunct{\mcitedefaultmidpunct}
{}{\mcitedefaultseppunct}\relax
\EndOfBibitem
\bibitem[Not()]{Note-1}
It should be noted that in a previous study, no ferromagnetism was observed in
  M=8 insulating superlattices\cite{Grutter2013} A careful comparison of the
  transport data of the M=8 superlattices in this current study to those of the
  previous study reveals that the present SLs have higher resistivity at a
  given N and M. These differences in resistivity suggest there may be
  differences in oxygen stoichiometry and cation oxidation states at the
  interface between sample sets. These differences are plausible given the
  different growth conditions\relax
\mciteBstWouldAddEndPuncttrue
\mciteSetBstMidEndSepPunct{\mcitedefaultmidpunct}
{\mcitedefaultendpunct}{\mcitedefaultseppunct}\relax
\EndOfBibitem
\bibitem[May \latin{et~al.}(2010)May, Kim, Rondinelli, Karapetrova, Spaldin,
  Bhattacharya, and Ryan]{May2010}
May,~S.~J.; Kim,~J.-W.; Rondinelli,~J.~M.; Karapetrova,~E.; Spaldin,~N.~A.;
  Bhattacharya,~A.; Ryan,~P.~J. \emph{Phys. Rev. B} \textbf{2010}, \emph{82},
  014110\relax
\mciteBstWouldAddEndPuncttrue
\mciteSetBstMidEndSepPunct{\mcitedefaultmidpunct}
{\mcitedefaultendpunct}{\mcitedefaultseppunct}\relax
\EndOfBibitem
\bibitem[Aschauer \latin{et~al.}(2013)Aschauer, Pfenninger, Selbach, Grande,
  and Spaldin]{Aschauer2013}
Aschauer,~U.; Pfenninger,~R.; Selbach,~S.~M.; Grande,~T.; Spaldin,~N.~A.
  \emph{Phys. Rev. B} \textbf{2013}, \emph{88}, 054111\relax
\mciteBstWouldAddEndPuncttrue
\mciteSetBstMidEndSepPunct{\mcitedefaultmidpunct}
{\mcitedefaultendpunct}{\mcitedefaultseppunct}\relax
\EndOfBibitem
\end{mcitethebibliography}

\end{document}